\documentclass{article}  
\usepackage{spadre2008}
\usepackage{graphicx}
\usepackage{horowitz}
\usepackage{fslash}

\newcommand{\wv}[1]{\mathbf{#1}}

\newcommand{\mg}{m_\gamma}

\newcommand{\mmq}{M_Q}

\frompage{000} \topage{000}                                              
\def\rightmark{Heavy Quark Photon Bremsstrahlung}
\def\leftmark{W.\ A.\ Horowitz}
 
\title{\zeroth Order Heavy Quark Photon Bremsstrahlung} 
\authors{
{W.\ A.\ Horowitz$^{1,a}$ %
}\\[2.812mm]
{\normalsize
\hspace*{-8pt}$^1$ Department of Physics, Columbia University, \\ 
538 W.\ 120$^\mathrm{th}$ St.\, New York, NY 10027, USA\\[0.2ex] 
}}
 
\abstract{We calculate the \zeroth order in opacity number distribution of massive photons (gluons) for heavy quark production radiation including interference from the away-side jet.  While consistent with the soft photon (gluon) approximation, we find that taking $1-x\approx 1$, as done in previous calculations, strongly affects the magnitude of energy loss.  Restoring gauge invariance by including the radiation associated with the away-side jet fills in the ``dead cone,'' but is a relatively small effect.  The Ter-Mikayelian reduction from vacuum energy loss is 10-40\% for 5-25 GeV charm and bottom quarks.}

 
\begin{document}
 
\maketitle
\setcounter{page}{1}

\begin{notes}
\item[a]
E-mail: horowitz@phys.columbia.edu
\end{notes}

\section{Introduction}\label{intro}
%
%
%
%
%

The confidence in the application of perturbative QCD methods to jet energy loss in heavy ion collisions gained from the early quantitative understanding of $\pi$ and $\eta$ suppression with null direct-$\gamma$ control \cite{Akiba:2005bs,Vitev:2002pf} has recently come into serious doubt \cite{ColeTalk}.  Evidence from measurements such as \highpt correlations \cite{Ackermann:2000tr,Winter:2006iz} and nonphotonic electrons \cite{Adare:2006nq,Abelev:2006db} demonstrates clear disagreement with perturbative models \cite{Shuryak:2001me,Horowitz:2005ja,Djordjevic:2005db,Wicks:2005gt,Armesto:2005mz}.  Several papers postulate alternative nonperturbative energy loss mechanisms \cite{van Hees:2005wb,Adil:2006ra,Gubser:2006bz,Herzog:2006gh}, and a new measurement, the double ratio of charm to bottom nuclear modification factors, has been suggested as a robust observable for testing some of these novel ideas \cite{Horowitz:2007su}.  Electromagnetic radiation from quark jets, as it is transparent to the partonic medium, holds enormous promise as a new tool for investigating jet energy loss mechanisms; naively one expects the spectrum of photons emitted from a jet that underwent a smooth, exponential slowdown will differ greatly from one that suffered the emission of a few hard gluons will differ from one that experienced a large number of soft scatterings.  Nascent data of $\gamma$-hadron correlations from $p+p$ collisions additionally motivates the theoretical exploration of photon bremsstrahlung in heavy ion collisions \cite{Hanks:2007qx}.  

In this paper we calculate as a warmup problem, and ultimately as an interesting problem in its own right, the \zeroth order in opacity energy loss of a heavy quark jet, the radiation associated with the production of a hard parton.  We will generalize the problem of a zero mass quark emitting a zero mass photon to a massive quark emitting a massive photon.  The lack of theoretical consistency in the understanding of light and heavy flavor jet suppression makes massive quark calculations of especial interest \cite{Horowitz:2005ja,Djordjevic:2005db,Wicks:2005gt}; moreover heavy quark predictions from pQCD will be necessary for comparison to AdS/CFT heavy quark drag results \cite{Horowitz:2007su}.  
For the case of \zeroth order emission QCD and QED are identical but for the replacement of $\alpha_{EM}$ with $\alpha_s$ and a color Casimir.  Using a massive photon will allow a comparison to already published results on the QCD Ter-Mikayelian effect \cite{Djordjevic:2003be}, whose main results were: (1) the Ter-Mikayelian effect leads to a large reduction in \zeroth order energy loss ($\sim30\%$ for charm quarks), (2) the full 1-loop HTL gluon propagator can be well approximated by using a fixed gluon mass $m_g=m_\infinity=\mu/\sqrt{2}$, and (3) the small-$x$, soft gluon, number distribution for \zeroth order in opacity is
\be
\label{magda}
\frac{dN_\mathrm{pQCD}^{(0)}}{d^3k} = \frac{Q^2\alpha}{\pi^2\omega} \frac{\wv{k}^2}{[\wv{k}^2+\mg^2+x^2\mmq^2]^2},
\ee
where, as usual, a bold variable represents a transverse two-vector.

\section{Calculation}
As a first step to compare to previously published results \cite{Djordjevic:2003be} we found the number distribution of emitted photons when simply plugging in massive 4-vectors into a standard classical E\&M calculation:
\bea
\label{eem}
E_\mathrm{EM} & = & \int\frac{d^3k}{(2\pi)^3}\frac{e^2}{2}\sum_{\lambda=1,2}\left|\vec{\epsilon}_\lambda(\vec{k})\cdot\left( \frac{\vec{p}'}{k\cdot p'} - \frac{\vec{p}}{k\cdot p} \right)\right|^2 \\
\label{preem}
& = & \int\frac{d^3k}{(2\pi)^3}\frac{e^2}{2} \left( \frac{2p\cdot p'}{(k\cdot p')(k\cdot p)}-\frac{\mmq^2}{(k\cdot p')^2}-\frac{\mmq^2}{(k\cdot p)^2} \right) \\
\label{em}
\Rightarrow \frac{dN_\mathrm{E\&M}^{(0)}}{d^3k} & = & \frac{Q^2\alpha}{\pi^2\omega} (1-x)^2 \frac{ \wv{k}^2 + (1-x)^2 \mg^2 }{[ \wv{k}^2 + (1-x)^2 \mg^2 + x^2\mmq^2 ]^2}+\mathcal{O}\left(1/E^+\right),
\eea
where to get from the first to the second line we used the completeness relation $\sum_{\lambda=1,2}\epsilon_\lambda^\mu\epsilon_\lambda^{\nu*}\rightarrow -g_{\mu\nu}$, and where we took
\bea
p & = & [(1-x)E^+,\frac{\mmq^2+\wv{k}^2}{(1-x)E^+},-\wv{k}] \\
p' & = & [\frac{\mmq}{E^+},E^+,0] \\
k & = & [xE^+,\frac{\mg^2+\wv{k}^2}{xE^+},\wv{k}],
\eea
with brackets indicating lightcone coordinates.  

There are two elements seen in \eq{em} and not in  \eq{magda}: several factors of $(1-x)^2$, and an $\mg^2$ in the numerator.  
The first makes no difference in the limit of small $x$; however energy loss calculations integrate over all $x$, and it turns out that neglecting these factors is a large effect.  The second simply cannot be reconciled with \eq{magda}.  Interestingly this extra mass term in the numerator fills in the ``dead cone,'' the region of small angles with respect to the jet axis for which $dN_g/dx\rightarrow0$ in \eq{magda} as $\wv{k}\rightarrow0$ when $\mmq\ne0$; this motivates additional study because naively the dead cone leads to a reduction in heavy quark energy loss, inconsistent with the observation of similar suppression patterns for pions, decay fragments from gluons and light quarks, and nonphotonic electrons, decay fragments from heavy charm and bottom quarks.



One may rightly object that the results of \eq{em} were derived using the usual massless photon E\&M formulae.  Surprisingly the only modification of \eq{eem} when using the Proca Lagrangian is to change the polarization sum to include the longitudinal mode.  It turns out that the extra terms generated by the application of the identity $\sum_{\lambda=1,2,3}\epsilon_\lambda^\mu\epsilon_\lambda^{\nu*}= -g_{\mu\nu}+k_\mu k_\nu/\mg^2$ exactly cancel, and \eq{em} is also valid for massive photon radiation.

\bfig[!ht]
\begin{center}
\leavevmode
\includegraphics[width=.75 \columnwidth]{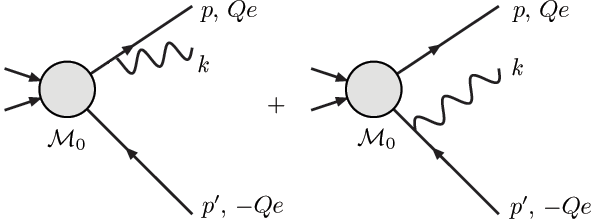}
\caption{\label{diagrams}\captionsize{The two diagrams contributing to the \zeroth order in opacity photon/gluon radiation spectrum.  Note the inclusion of the radiation from the away side jet, which is usually ignored in pQCD calculations.}}
\end{center}
\efig

In order to understand the discrepancy from the field theory perspective, consider the diagrams contributing to the \zeroth order shown in \fig{diagrams}.  Evaluation of these leads to
\be
\label{feyn}
i\mathcal{M} = Qe \bar{u}(p)\left[ \frac{2p\cdot\epsilon^*+\FMslash{\epsilon}^*\FMslash{k}}{2p\cdot k + m_\gamma^2}\mathcal{M}_0 - \mathcal{M}_0 \frac{2p'\cdot\epsilon^*+\FMslash{\epsilon}^*\FMslash{k}}{2p'\cdot k + m_\gamma^2} \right] v(p'),
\ee
where we have taken $\mathcal{M}_0(p+k,p')\approx\mathcal{M}_0(p,p'+k)\approx\mathcal{M}_0$ in the small $x$, soft radiation limit.  Most pQCD calculations ignore the away-side jet; one can easily see that the second term in \eq{feyn}, corresponding to the inclusion of the second diagram in \fig{diagrams}, is crucial for preserving the Ward identity \cite{Wang:1994fx}.  Simultaneously dropping the $\FMslash{k}$ in the numerator and the $m_\gamma$ in the denominator (consistent with the soft photon limit) exactly reproduces the classical Proca result.  We note that assuming $\mathcal{M}_0$ commutes with $\FMslash{\epsilon}^*\FMslash{k}$ and retaining $\mg\ne0$ in the denominator of \eq{feyn} results in a $dN_\mathrm{pQCD}/d^3k$ with leading order identical to \eq{em} but with $(1-x)^2\rightarrow(1-x/2)^2$ as the prefactor of the $\mg$ in the numerator and $(1-x)^2\rightarrow(1-x)$ as the prefactor of the $\mg$ in the denominator. 

\section{Size of Effects}
\begin{figure}[htb!]
\begin{center}
\leavevmode
$\begin{array}{c@{\hspace{.00in}}c}
\includegraphics[width=.5\columnwidth]{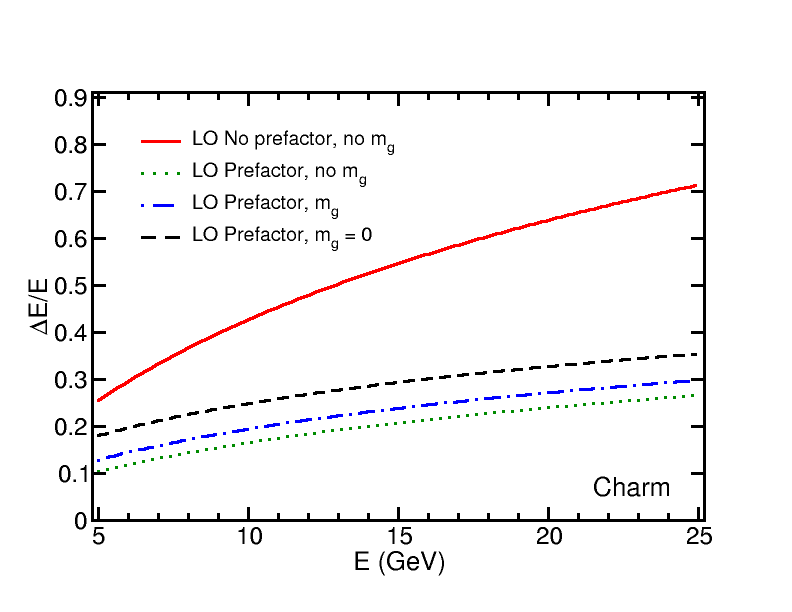} & 
\includegraphics[width=.5\columnwidth]{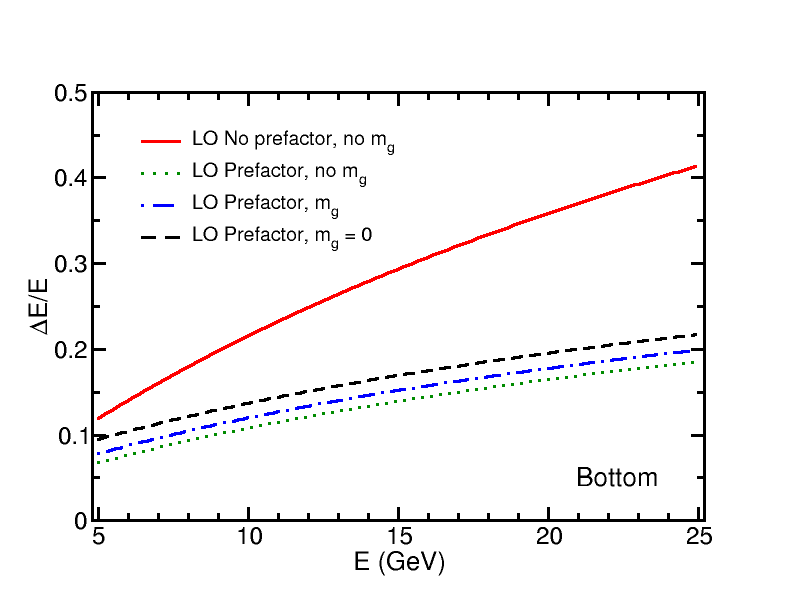} \\ [-0.in]
{\mbox {\scriptsize {\bf (a)}}} & {\mbox {\scriptsize {\bf (b)}}}
\end{array}$
\vspace{-.15in}
\caption{
\label{effectsone}\captionsize{(Color online) \zeroth order radiative energy loss for (a) charm and (b) bottom quarks.  All results are to leader order (LO) in $1/E^+$.  One sees that the largest effect (50-150\%) comes from including the $(1-x)^2$ prefactor and that filling in the ``dead cone'' with the massive photon is a rather small one (5-20\%).  Comparison with $\mg=0$ yields the magnitude of the LO Ter-Mikayelian effect (10-40\%).}
}
\end{center}
\vspace{-.2in}
\end{figure}

We wish to investigate quantitatively the effect of these extra terms on the \zeroth order energy loss.  To do so we enforce physicality by restricting the $x$ and $\wv{k}$ integration limits so that the emitted photon has $E_\gamma \gte m_g$ and leaves the jet with $E_{jet}\gte\mmq$.  For ease of comparison with \cite{Djordjevic:2003be} we set $\mu=.5$ GeV and $\alpha=.5$ fixed.  One can see from \fig{effectsone} the large (50-150\%) effect on $\Delta E/E$ of including the overall prefactor of $(1-x)^2$.  
Filling in the ``dead cone'' makes only a small difference to the energy lost (5-20\%); this is a surprise as the ``dead cone'' is the usual naive justification for heavy quarks having smaller radiative energy loss than light quarks.

\begin{figure}[htb!]
\begin{center}
\leavevmode
$\begin{array}{c@{\hspace{.00in}}c}
\includegraphics[width=.5\columnwidth]{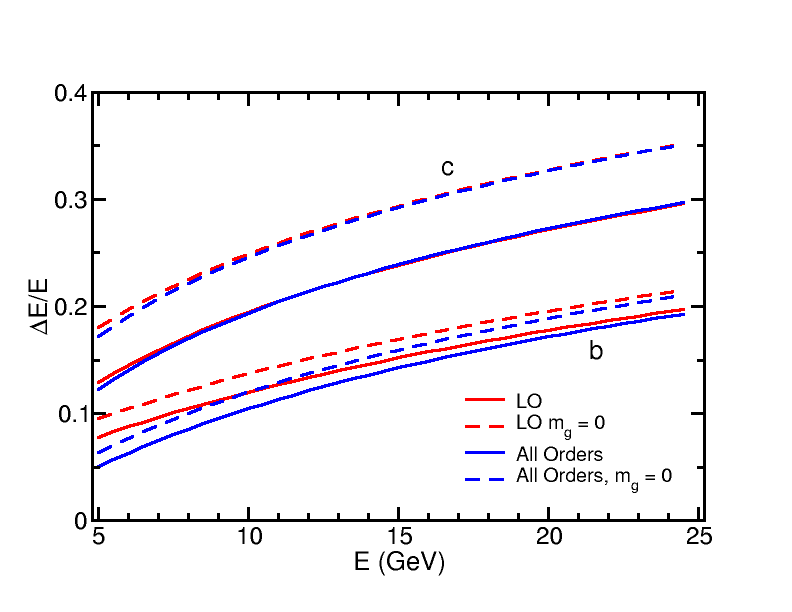} & 
\includegraphics[width=.5\columnwidth]{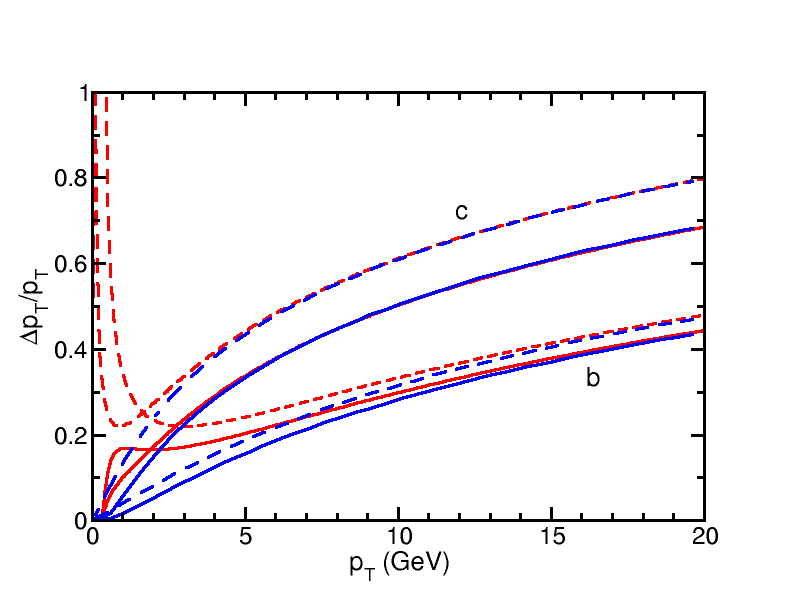} \\ [-0.in]
{\mbox {\scriptsize {\bf (a)}}} & {\mbox {\scriptsize {\bf (b)}}}
\end{array}$
\vspace{-.15in}
\caption{
\label{effectstwo}\captionsize{The effect of including all terms from \eq{em} instead of just the leading order (LO) terms in $1/E^+$ for (a) $\Delta E/E$ and (b) $\Delta\eqnpt/\eqnpt$ (the legend in (a) applies to both plots).  For $\Delta E/E$, the size of the relative difference in magnitude--the Ter-Mikayelian effect--is changed little while the overall normalization is significantly altered at low energies.  For $\Delta\eqnpt/\eqnpt$ both the relative and overall normalizations change quite a bit, with the inclusion of all terms regulating the $\eqnpt\rightarrow0$ divergences in the vacuum production radiation spectrum.
}
}
\end{center}
\vspace{-.2in}
\end{figure}

\fig{effectstwo} demonstrates the effect of including all the terms generated by \eq{em}, not just the LO in $1/E^+$.  Of course at higher $E$ and $\eqnpt$ the additional terms make little difference, but they regulate the otherwise divergent results in $\Delta \eqnpt/\eqnpt$ as $\eqnpt\rightarrow0$.  The Ter-Mikayelian effect, given by the difference between the $\mg\ne0$ and $\mg=0$ plots in \fig{effectstwo}, varies from 10-40\% for charm and bottom energy loss.

\section{Conclusions}\label{concl}
Unfortunately after many years of effort there is still no single satisfactory energy loss model for heavy ion collisions at RHIC.  This leads to the need for basic experimental tests of the gross features of the underlying energy loss mechanism, whether it be more like pQCD, AdS/CFT, or some other approximation.  
Medium induced photon bremsstrahlung has the potential to provide unprecedented insight into the modes of jet energy loss, and in this paper we took the intermediate step of analyzing the \zeroth order in opacity production radiation energy loss.  
While enforcing gauge invariance by not neglecting the away side jet fills in the ``dead cone,'' this ultimately has only a small effect on the radiation spectrum.  On the other hand neglecting the overall factor of $(1-x)^2$ in the emitted photon distribution makes a surprisingly large difference.  This prefactor, also neglected in medium-induced gluon radiation derivations \cite{Gyulassy:2000er,Djordjevic:2003zk}, may significantly alter \raapt calculations, especially at smaller momenta.

 
 


\vfill\eject
\end{document}